\documentclass[journal=jacsat,manuscript=article]{achemso}

\usepackage{multicol}
\usepackage[version=4]{mhchem}
\usepackage{graphicx}% Include figure files
\usepackage{amsmath,amsfonts,amsthm,bm}
\usepackage{subcaption}
              
\usepackage{hyperref}
\usepackage{physics}

\usepackage{nameref}
\usepackage{zref-xr}
\usepackage{zref-user}
\zxrsetup{toltxlabel}

\author{Alexis Ralli}
\affiliation{Centre for Computational Science, Department of Chemistry, University College London, WC1H 0AJ, United Kingdom}
\email{alexis.ralli.18@ucl.ac.uk}
\author{Michael I. Williams de la Bastida}
\affiliation{Centre for Computational Science, Department of Chemistry, University College London, WC1H 0AJ, United Kingdom}
\email{michael.williams.20@ucl.ac.uk}
\author{ Peter\ V.\ Coveney}
\affiliation{Centre for Computational Science, Department of Chemistry, University College London, WC1H 0AJ, United Kingdom}
\alsoaffiliation{
UCL Centre for Advanced Research Computing, Gower Street , London
WC1E 6BT, United Kingdom}%
\alsoaffiliation{
Informatics Institute, University of Amsterdam, Amsterdam, 1098 XH, Netherlands
}%
\email{p.v.coveney@ucl.ac.uk}

\title[An \textsf{achemso} demo]
  {A Scalable Approach to Quantum Simulation via Projection-Based Embedding}

\abbreviations{VQE,QPE,DFT}

\begin{document}

% \begin{tocentry}

% Some journals require a graphical entry for the Table of Contents.
% This should be laid out ``print ready'' so that the sizing of the

% Inside the \texttt{tocentry} environment, the font used is Helvetica
% 8\,pt, as required by \emph{Journal of the American Chemical

% The surrounding frame is 9\,cm by 3.5\,cm, which is the maximum
% permitted for  \emph{Journal of the American Chemical Society}
% graphical table of content entries. The box will not resize if the
% content is too big: instead it will overflow the edge of the box.

% This box and the associated title will always be printed on a
% separate page at the end of the document.

% \end{tocentry}

\begin{abstract}
Owing to the computational complexity of electronic structure algorithms running on classical digital computers, the range of molecular systems amenable to simulation remains tightly circumscribed even after many decades of work. Quantum computers hold the promise of transcending such limitations although in the current era the size and noise of these devices militates against significant progress. Here we describe a new and chemically intuitive approach that permits a subdomain of a molecule’s electronic structure to be calculated accurately on a quantum device, while the rest of the molecule is described at a lower level of accuracy using density functional theory running on a classical computer. We demonstrate that our method produces improved results for molecules that cannot be simulated fully on quantum computers but which can be resolved classically at a lower level of approximation. Our algorithm is tunable, so that the size of the quantum simulation can be adjusted to run on available quantum resources. Therefore, as quantum devices become larger, our method will enable increasingly large subdomains to be studied accurately.
\end{abstract}

% \begin{multicols}{2}

\section{\zlabel{sec:intro}Introduction}
Quantum computing is anticipated to enable accurate simulation of chemical systems beyond the capabilities of classical methods. Whether this aim will be achieved with so-called Noisy Intermediate-scale Quantum (NISQ) processors, is still to be seen\cite{preskill2018quantum, cordier2021biology, Cheng2020, McArdle2020}. While devices are improving rapidly, NISQ applications also require algorithmic tools to mitigate noise and reduce required qubit counts.\

Embedding procedures work by first partitioning a system and then applying differing levels of theory to each region. An accurate but computationally expensive method is applied to a small \emph{active region}\cite{Sun2016, Bauer2020}. The surrounding \emph{environment} is handled with a more efficient but approximate method. This allows some of the physically relevant detail to be captured while avoiding the computational cost of accurately simulating the entire system. However, even for fairly small active regions, exact classical simulation using the Full Configuration Interaction (FCI) method quickly becomes unfeasible due to the number of Slater determinants (states) scaling factorially as $\binom{M}{N}$, for $N$ electrons and $M$ orbitals \cite{McArdle2020}.

The current ``gold standard" in conventional quantum chemistry is coupled cluster (CC) theory, which offers a good accuracy-to-cost ratio  and reduces this factorial complexity \cite{bartlett2007coupled, romero2018strategies}. The CC single double (CCSD) method scales as $\mathcal{O}(M^{6})$ \cite{purvis1982full}. The CCSD(T), which treats the triple excitations pertubatively, scales as $\mathcal{O}(M^{7})$ in time \cite{McArdle2020}.  This still imposes practical limitations on system size while imperfectly approximating the effects of correlation \cite{Herbert2019}. Therefore, classical embedding methods still inevitably inherit the shortcomings of such methods, even within a smaller active region. In short, accurately simulating quantum effects at large scale remains elusive.

Quantum computers can efficiently represent the state of general quantum systems and provide a practical way to perform quantum chemistry simulations in polynomial time \cite{aspuru2005simulated}. However, this approach will only be possible in the fault tolerant regime, as it requires the quantum phase estimation (QPE) algorithm \cite{kitaev1995quantum} which cannot be implemented on current NISQ quantum computers \cite{o2016scalable, mohammadbagherpoor2019experimental}. Quantum algorithms designed for NISQ devices, such as the variational quantum eigensolver (VQE) \cite{peruzzo2014variational}, allow quantum systems to be studied using present day hardware; however, this is limited by the current quality and quantity of qubits. To date, the largest chemical simulation was a $12$ qubit VQE study of an \ce{H12} chain \cite{google2020hartree}.

By embedding a wave function simulation calculated on a quantum computer into a larger classical simulation, we can mitigate some of the shortfalls of classical hardware in describing quantum systems, while requiring fewer qubits and shorter quantum circuits than full-system quantum simulation. This will allow systems normally too large to study at the wave function level to be modelled via a multi-scale approach. In this way, embedding can serve as an algorithmic tool to mitigate the shortcomings of quantum and classical processors, thereby providing novel results. Additionally, as embedding methods may utilise fault-tolerant quantum simulation methods, they will continue to facilitate the study of systems larger than would otherwise be possible. Hybrid embedding methods published to date include wave function-in-DFT \cite{Rossmannek2020, Ma2020}, Density Matrix Embedding Theory (DMET) \cite{Rubin2016, Yamazaki2018} and Dynamical Mean-Field Theory (DMFT) \cite{Bauer2016, Kreula2016, Steckmann2021} approaches.

We present a projection-based embedding method which enables the application of quantum algorithms to molecules of arbitrary size while consistently improving on the results of full-system Density Functional Theory (DFT). This method outputs a Hamiltonian which can be solved using any suitable NISQ or fault-tolerant quantum algorithm, thus augmenting the usefulness of quantum processors in general. We anticipate that by targeting quantum processors at regions with strong correlation, hybrid embedding will enable novel results.

\section{Projection Based Embedding}

Projection-based embedding, which was first proposed by Manby \textit{et al.} \cite{Manby2012}, provides a practical way to perform formally exact quantum embedding. We summarise the important details of their approach here.

To begin, an initial DFT calculation of the entire system is carried out using a low (cheap) level of theory. This yields a set of molecular orbitals (MOs) $\{ \psi_{i}(\vec{r}) | i=1,2,\hdots , N \}$. Each MO is formed from a linear combination of $K$ known atomic orbital (AO) basis functions $\{ \phi_{j}(\vec{r}) | j=1,2,\hdots , K\}$:

\begin{equation}
\zlabel{eq:MOs_def}
    \psi_{i}(\vec{r}) =  \sum_{j=1}^{K} \bm{C}_{j i} \phi_{j} (\vec{r}),
\end{equation}
where $\bm{C}$ is a matrix of MO coefficients. In general, the AO basis functions $\phi_{j}(\vec{r})$ are not orthonormal. However, linear combinations of these non-orthogonal AOs, given by the columns of $\bm{C}$, construct orthogonal MOs $\psi_{i}(\vec{r})$. %This can be seen from $\bm{C}^{T}\bm{S}\bm{C}=\bm{I}$, where $\vb{S}$ is the AO overlap matrix:
% \begin{equation}
%      \vb{S}_{\mu \nu} = \int d\vec{r} \; \phi_{\mu}(\vec{r})^{*}\phi_{\nu}(\vec{r}) = \bra{\phi_{\mu}}  \phi_{\nu}\rangle .
% \end{equation} 

% https://pubs.acs.org/doi/pdf/10.1021/ar500082t
% https://pubs.acs.org/doi/pdf/10.1021/ct400793q
Next we localize the canonical MOs $\psi_{i}$ via different localisation methods - described in further detail in the Supporting Information. In effect, we use a unitary transform $\bm{U}$ (defined by the localization procedure) to spatially localize each $\ket{\psi_{i}}$ as much as possible. We denote these orbitals as localized molecular orbitals (LMOs) or $\ket{\psi_{i}^{LMO}}$, which are defined as \cite{lehtola2013unitary}: %: $\ket{\psi_{i}^{LMO} (\vec{r})} = \ket{\psi_{i} (\vec{r})} U = \sum_{j=1}^{K} \bm{C}_{j i} U \ket{\phi_{j} (\vec{r})} = \sum_{j=1}^{K} \bm{C}_{ji}^{LMO} \ket{\phi_{j} (\vec{r})}$, where $\bm{C}^{LMO} = \bm{C}U$. In this work we localize only the occupied MOs.

% \begin{equation}
% \zlabel{eq:LMOs_def}
% \begin{aligned}
%     \ket{\psi_{i}(\vec{r})} \mapsto \sum_{j=1}^{K} \bm{U}_{ji}\ket{\psi_{i}^{LMO}(\vec{r})},
% \end{aligned}
% \end{equation}

\begin{equation}
\zlabel{eq:LMOs_def}
\begin{aligned}
    \ket{\psi_{i}^{LMO}} \mapsto \sum_{b=1}^{K} \bm{U}_{bi}\ket{\psi_{b}},
\end{aligned}
\end{equation}
where $\bm{U}\bm{U}^{\dagger}=\bm{U}^{\dagger}\bm{U}=\bm{I}$. We can write $\bm{C}$ under this transformation as:

\begin{equation}
\zlabel{eq:comb}
\begin{aligned}
    \ket{\psi_{i}^{LMO}} &= \bm{U}\ket{\psi_{i}} = \sum_{b=1}^{K} \bm{U}_{bi}\ket{\psi_{b}}  = \sum_{b=1}^{K} [\bm{U}^{T}]_{ib}\ket{\psi_{b}}  \\
    &= \sum_{b=1}^{K} [\bm{U}^{T}]_{ib} \Big(\sum_{j=1}^{K} \bm{C}_{j b} \ket{\phi_{j}} \Big)\\
    &=  \sum_{b=1}^{K} \sum_{j=1}^{K}  [\bm{U}^{T}]_{ib} \bm{C}_{j b} \ket{\phi_{b}} \\
    &= \sum_{j=1}^{K}  [\bm{C} \bm{U}^{T}]_{ji} \ket{\phi_{j}}\\
    &= \sum_{a=1}^{K}  \bm{C}_{ji}^{LMO} \ket{\phi_{j}}.
\end{aligned}
\end{equation}
Here $\bm{C}^{LMO} = \bm{C}\bm{U}^{T}$ and the columns of this matrix give each localized MO. In this work we localize only the occupied MOs and leave the virtual orbitals unchanged.
% \begin{equation}
% \zlabel{eq:LMOs_def}
% \begin{aligned}
%     \ket{\psi_{i}(\vec{r})} \mapsto \sum_{j} U_{ji}\ket{\psi_{i}(\vec{r})}  \\
% \end{aligned}
% \end{equation}

For a closed shell molecule, described by a single determinant wave function, each MO $\psi_{i}$ contains two electrons and thus the total charge density is \cite{szabo2012modern}:

\begin{equation}
\zlabel{eq:charge_denistry}
\begin{aligned}
    \rho(\vec{r}) &=  2\sum_{i=1}^{N/2} \psi_{i}^{*}(\vec{r}) \psi_{i}(\vec{r}) \\
    &= 2\sum_{i=1}^{N/2} \Bigg( \underbrace{\sum_{\nu=1}^{K} \bm{C}_{\nu i}^{*} \phi_{\nu}^{*} (\vec{r})}_{\psi_{i}^{*}(\vec{r})} \underbrace{\sum_{\mu=1}^{K}  \bm{C}_{\mu i} \phi_{\mu} (\vec{r})}_{\psi_{i}(\vec{r})} \Bigg) \\
    &= \sum_{\mu=1}^{K} \sum_{\nu=1}^{K} \Big[ 2\sum_{i=1}^{N/2} \bm{C}_{\mu i} \bm{C}_{\nu i}^{*} \Big] \phi_{\nu}^{*}(\vec{r}) \phi_{\mu}(\vec{r}) \\
    &= \sum_{\mu=1}^{K} \sum_{\nu=1}^{K} \gamma_{\mu \nu} \phi_{\nu}^{*}(\vec{r}) \phi_{\mu}(\vec{r}). 
\end{aligned}
\end{equation}
Here the square brackets define the density matrix $\gamma_{\mu \nu}$ (defined in the AO basis):

\begin{equation}
\zlabel{eq:density_mat}
    \gamma_{\mu \nu} = 2\sum_{i=1}^{N/2} \bm{C}_{\mu i} \bm{C}_{\nu i}^{\dagger},
\end{equation}
that for a set of basis function $\{ \phi_{j}(\vec{r}) | j=1,2,\hdots , K\}$ fully specifies the charge density $\rho(\vec{r})$ \cite{szabo2012modern}. The sum runs over $N/2$, as these are the occupied MOs of a closed shell calculation. The whole matrix can be obtained as $\gamma^{total}= 2\bm{C}_{occ} (\bm{C}_{occ})^{\dagger}$, where $occ$ denotes only using the occupied columns of the $\bm{C}$ matrix (the first $N/2$ columns, indexed by $i$ in Equation \zref{eq:density_mat}). In the localized basis, the density matrix remains unchanged as:

\begin{equation}
\zlabel{eq:density_mat_loc}
\begin{aligned}
    \gamma^{total} &= 2\bm{C}_{occ} (\bm{C}_{occ})^{\dagger} \\
    &= 2[\bm{C}_{occ}^{LMO} \bm{U}] \; [\bm{U}^{\dagger}(\bm{C}_{occ}^{LMO})^{\dagger}]\\
    &= 2 \bm{C}_{occ}^{LMO} (\bm{C}_{occ}^{LMO})^{\dagger}.
\end{aligned}
\end{equation}
% $\gamma^{total}= 2\bm{C}_{occ} (\bm{C}_{occ})^{\dagger} = 2\bm{C}_{occ}^{LMO} (\bm{C}_{occ}^{LMO})^{\dagger}$, where $occ$ denotes only using the occupied columns of the whole matrix (the first $N/2$ columns, indexed by $i$ in the Equation \zref{eq:density_mat_loc}). 

%Note when we localize the molecular orbitals the charge density remains the unchanged:
% \begin{equation}
% \zlabel{eq:density_mat_loc}
% \begin{aligned}
%     \gamma_{\mu \nu} &= 2\sum_{i=1}^{N/2} \bm{C}_{\mu i} \bm{C}_{\nu i}^{*} \\
%         &=2\sum_{i=1}^{N/2} \bm{C}_{\mu i} \bm{U}_{\mu i}^{\dagger}\bm{U}_{\nu i} \bm{C}^{*}_{\nu i}\\
%     &=2\sum_{i=1}^{N/2} (\bm{C}\bm{U}^{\dagger})_{\mu i} (\bm{U}\bm{C}^{*})_{\nu i}\\
%     &=2\sum_{i=1}^{N/2} \bm{C}_{\mu i}^{LMO} \bm{C}_{\nu i}^{*,LMO}\\
% \end{aligned}
% \end{equation}
% The whole matrix can be obtained as $\gamma^{total}= 2\bm{C}_{occ} (\bm{C}_{occ})^{\dagger} = 2\bm{C}_{occ}^{LMO} (\bm{C}_{occ}^{LMO})^{\dagger}$, where $occ$ denotes only using the occupied columns of the whole matrix (the first $N/2$ columns, indexed by $i$ in the Equation \zref{eq:density_mat_loc}).

Given a set of localised molecular orbitals, we partition them into two subsystems denoted $act$ (active) and $env$ (environment). There are different methods to do so and we summarise our approach in the Supporting Information. Overall we generate a set of (occupied) LMO indices $\mathcal{K}$ and $\mathcal{L}$ for the active and environment subsystems respectively. The resulting charge density for each subsystem can then be written as:

% https://web.northeastern.edu/afeiguin/phys5870/phys5870/node19.html
\begin{subequations}
    \begin{equation}
    \zlabel{eq:dm_act}
       \gamma_{\mu \nu}^{act} = 2 \sum_{k \in \mathcal{K}} \bm{C}_{\mu k}^{LMO} (\bm{C}_{\nu k}^{LMO})^{\dagger},
    \end{equation}
\begin{equation}
    \zlabel{eq:dm_env}
       \gamma_{\mu \nu}^{env} = 2 \sum_{l \in \mathcal{L}} \bm{C}_{\mu l}^{LMO} (\bm{C}_{\nu l}^{LMO})^{\dagger},
\end{equation}
\end{subequations}
for closed-shell calculations. The set $\mathcal{K} \cup \mathcal{L}$ contains all the occupied molecular orbital indices.

% \begin{subequations}
%     \begin{equation}
%     \zlabel{eq:dm_act}
%       \gamma^{act} = 2 \sum_{k \in \mathcal{K}} \ket{\psi_{k}^{LMO}} \bra{\psi_{k}^{LMO}},
%     \end{equation}
%     %%
% \begin{equation}
%     \zlabel{eq:dm_act_terms}
%       \gamma^{act}_{pq} = 2 \sum_{k \in \mathcal{K}} C_{pk}^{LMO} C_{qk}^{LMO *}.
%     \end{equation}
% \end{subequations}
% and
% \begin{subequations}
%     \begin{equation}
%     \zlabel{eq:dm_env}
%       \gamma^{env} = 2 \sum_{l \in \mathcal{L}} \ket{\psi_{l}^{LMO}} \bra{\psi_{l}^{LMO}},
%     \end{equation}
%     %%
% \begin{equation}
%     \zlabel{eq:dm_env_terms}
%       \gamma_{pq}^{env} = 2 \sum_{l \in \mathcal{L}} C_{pl}^{LMO} C_{ql}^{LMO*}.
%     \end{equation}
% \end{subequations}
% for closed-shell calculations.

The total system electron density is written as a sum of subsystem densities:
\begin{equation}
\zlabel{eq:total_density}
\begin{aligned}
    \gamma^{total} &=  \gamma^{act} + \gamma^{env} \\
    &= 2\bm{C}_{\mathcal{K}}^{LMO} (\bm{C}_{\mathcal{K}}^{LMO})^{\dagger} + 2\bm{C}_{\mathcal{L}}^{LMO} (\bm{C}_{\mathcal{L}}^{LMO})^{\dagger} \\
    &= 2\bm{C}_{occ}^{LMO} (\bm{C}_{occ}^{LMO})^{\dagger}.
\end{aligned}
\end{equation}
The number of electrons will also be split according to $n_{e}^{total} = n_{e}^{act} + n_{e}^{env} = tr(\vb{S}\gamma^{act}) + tr(\vb{S}\gamma^{env})=tr(\vb{S}\gamma^{total})$, where $tr$ denotes the trace operation and $\vb{S}$ is the AO overlap matrix:

\begin{equation}
     \vb{S}_{\mu \nu} = \bra{\phi_{\mu}}  \phi_{\nu}\rangle = \int d\vec{r} \; \phi_{\mu}(\vec{r})^{*}\phi_{\nu}(\vec{r}).
\end{equation}

The energy of the full system can be found from its components via \cite{Claudino2019}:

\begin{equation}
\zlabel{eq:full_energy}
\begin{aligned}
    E[\gamma^{act}, \gamma^{env}] =  &\underbrace{tr(\gamma^{act} \vb{h}_{core}) + \vb{g}(\gamma^{act})}_{\text{energy of isolated $act$ system}} + \\ &\underbrace{tr(\gamma^{env} \vb{h}_{core}) + \vb{g}(\gamma^{env})}_{\text{energy of isolated $env$ system}}+ \\
    &\underbrace{\vb{g}(\gamma^{act}, \gamma^{env})}_{\text{nonadditive two-electron energy}}.
\end{aligned}
\end{equation}
Here $\vb{h}_{core}$ is the one-electron core Hamiltonian and $\vb{g}$ groups the two-electron terms - Coulomb and exchange for Hartree-Fock and exchange-correlation for DFT. The nonadditive two-electron energy is given by:

\begin{equation}
   \vb{g}(\gamma^{act}, \gamma^{env}) =\vb{g}(\gamma^{act} + \gamma^{env})  - \vb{g}(\gamma^{act}) - \vb{g}(\gamma^{env}),
\end{equation}
and accounts for the interaction between subsystems \cite{Claudino2019}.

Next we want to solve the active system using a higher (more accurate) level of theory. The effect of the interaction between the active and environment subsystems is accounted for by additional terms in the core Hamiltonian. The Fock matrix for the active system embedded in the environment system is \cite{Manby2012}:

\begin{equation}
\zlabel{eq:emb_fock}
    \begin{aligned}
       \vb{F}_{emb}^{act} &= \vb{h}_{core} + \vb{V}_{emb}+ \vb{P}_{proj}^{env} + \vb{g}(\gamma_{emb}^{act}) \\
       &= \vb{h}_{emb} + \vb{g}(\gamma_{emb}^{act}),
    \end{aligned}
\end{equation}

\noindent where:

\begin{subequations}
    \begin{equation}
    \zlabel{eq:V_emb}
       \vb{V}_{emb} = \vb{g}(\gamma^{act} + \gamma^{env}) - \vb{g}(\gamma^{act}),
    \end{equation}
\begin{equation}
    \zlabel{eq:H_emb}
       \vb{h}_{emb} =  \vb{h}_{core} + \vb{V}_{emb}+ \vb{P}_{proj}^{env}.
    \end{equation}
\end{subequations}

% \begin{equation}
% \zlabel{eq:V_emb}
%   \vb{V}_{emb} = \vb{g}(\gamma^{act} + \gamma^{env}) - \vb{g}(\gamma^{act}).
% \end{equation}

The embedding potential $\vb{V}_{emb}$ describes all the interactions (nonadditive part) between the active and environment subsystems \cite{lee2019projection}. Due to the subsystem densities (Equation \zref{eq:total_density}) being constructed from disjoint subsets of orthogonal orbitals, the normally difficult-to-evalute nonadditive kinetic potential (NAKP) terms \cite{roncero2008inversion} are exactly zero \cite{Manby2012,barnes2013accurate, lee2019projection}.

$\vb{P}_{proj}^{env}$ is a projection operator that enforces inter-subsystem (orbital) orthogonality. There are different ways to define this operator and we consider two in this work. The first definition was proposed by the Manby and Miller groups\cite{Manby2012}. They use a parameter ($\mu$) to shift the orbital energies of the environment to high energies - effectively meaning they will never by occupied. This projector is defined as:

% \begin{equation}
%   P_{proj}^{env} = 
%     \begin{cases}
%       \mu (\vb{S} \gamma^{env} \vb{S}) & \text{if $proj=mu$}\\
%       -\frac{1}{2}\big( \vb{F} \gamma^{env} \vb{S} + \vb{S} \gamma^{env} \vb{F} \big)  & \text{if $proj=huz$}
%     \end{cases}   
% \end{equation}

\begin{equation}
\zlabel{eq:mu_shift_eqation}
\begin{aligned}
   (\vb{P}_{\mu}^{env})_{ij} &= \mu \bra{\psi_{i}^{LMO}} \vb{P}^{env} \ket{\psi_{j}^{LMO}}  \\
     &= \mu [\vb{S} \gamma^{env} \vb{S}]_{ij},
\end{aligned}
\end{equation}
where $\mu$ is some large integer, $\vb{S}$ is the AO overlap matrix. $\vb{P}^{env}$ is a projector defined as: 

\begin{equation}
    \begin{aligned}
         \vb{P}^{env} = \sum_{l \in \mathcal{L}} \ket{\psi_{l}^{LMO}} \bra{\psi_{l}^{LMO}}.
    \end{aligned}
\end{equation}

Here we use the notation $l \in \mathcal{L}$ to mean the sum over the set of occupied MO indices for the environment orbitals. The work in \cite{Manby2012, Claudino2019} shows $\mu$ is numerically robust and can usually be set to $\mu=10^{6}$. In the limit that $\mu\to\infty$ this method is exact. The action of this operator with the Fock matrix is:

\begin{subequations}
    \begin{equation}
        \begin{aligned}
             (\vb{F} + \vb{P}_{\mu}^{env}) \ket{\psi_{k}^{LMO}} =  \epsilon_{k}^{act}  \ket{\psi_{k}^{LMO}},
        \end{aligned}
    \end{equation}
        \begin{equation}
        \zlabel{eq:mu_env_example}
        \begin{aligned}
              (\vb{F} + \vb{P}_{\mu}^{env}) \ket{\psi_{l}^{LMO}} =  (\epsilon_{l}^{env} + \mu )  \ket{\psi_{l}^{LMO}} \approx + \mu \ket{\psi_{l}^{LMO}}.
        \end{aligned}
    \end{equation}
\end{subequations}
Again, $k$ and $l$ represent occupied LMOs of the active and environment subsystems respectively. Qualitatively the orbital energies of the active system are left unchanged and the orbitals for the environment are pushed to very high energies as $\mu >> \epsilon_{i}^{env}$ - effectively suppressing transitions to these states and stopping hybridisation.

The second approach, proposed by Kallay \textit{et al.} \cite{hegely2016exact}, is to use the Huzinaga projector \cite{huzinaga1971theory, francisco1992generalized}:

\begin{equation}
\begin{aligned}
\zlabel{eq:huz_definition}
     \vb{P}_{huz}^{env} &= -\big( \vb{F} \vb{P}^{env} + \vb{P}^{env} \vb{F}  \big) \\
     &= -\frac{1}{2}\big( \vb{F} \gamma^{env} \vb{S} + \vb{S} \gamma^{env} \vb{F} \big).
\end{aligned}
\end{equation}
Note that the $-\frac{1}{2}$ prefactor is needed for closed-shell systems. This operator enforces orthogonality of the occupied orbitals of each subsystem \cite{Shimazaki2017}. The form of this operator increases the orbital energy for the occupied environment orbitals and leaves the active system unchanged. We write its action formally as:

\begin{subequations}
    \begin{equation}
        \begin{aligned}
             (\vb{F} + \vb{P}_{huz}^{env}) \ket{\psi_{k}^{LMO}} =  \epsilon_{k}^{act}  \ket{\psi_{k}^{LMO}},
        \end{aligned}
    \end{equation}
    \begin{equation}
    \zlabel{eq:huz_on_env}
        \begin{aligned}
              (\vb{F} + \vb{P}_{huz}^{env}) \ket{\psi_{l}^{LMO}} &= (\epsilon_{i}^{env} -2 \epsilon_{l}^{env})  \ket{\psi_{l}^{LMO}} \\
              &= -1 \epsilon_{l}^{env} \ket{\psi_{l}^{LMO}}.
        \end{aligned}
    \end{equation}
\end{subequations}
As $\epsilon_{l}^{env}$ for occupied orbitals should always be negative, this ensures the filled environment orbitals will always have a positive energy and thus will never be filled. Whereas in Equation \zref{eq:mu_env_example}, for the unlikely case that $\mu < \epsilon_{i}^{env}$,  the environment MOs  are not projected to high enough energies to stop hybridization. This scenario is highly improbable, but could still occur.

The Huzinaga formalism guarantees that $[\vb{P}_{huz}^{env}, \vb{F}_{emb}^{act}]=0$ and removes the need for the $\mu$ parameter shift \cite{Chulhai2017}. This methodology ensures that the environment orbitals $\ket{\psi_{l\in\mathcal{L}}^{LMO}}$ are eigenfunctions of $(\vb{F} + \vb{P}_{huz}^{env})$ and when solving the active system (Equation \zref{eq:emb_fock}) the resultant canonical active orbitals will be orthogonal to them. %Furthermore, it was shown in REF that these orbitals make equation X stationary with 

The energy of the active system embedded in the environment is given by:

\begin{equation}
\zlabel{eq:DFT_in_DFT}
\begin{aligned}
    E[\gamma_{emb}^{act} ; \gamma^{act}, \gamma^{env}] &= \mathcal{E}[\gamma_{emb}^{act}] + E[\gamma_{env}] + \vb{g}(\gamma^{act}, \gamma^{env})\\
     & + tr\Big( (\gamma_{emb}^{act}-\gamma^{act}) (\vb{V}_{emb}+ \vb{P}_{proj}^{env}) \Big),
\end{aligned}
\end{equation}
colloquially denoted as a DFT-in-DFT calculation.

We use the same notation as \cite{Claudino2019}, where $\mathcal{E}$ differs from $E$ as it allows for different functionals to be applied and is computed from the embedded density matrix of the active system. Note that Equation \zref{eq:emb_fock} is solved self-consistently to give $\gamma_{emb}^{act}$. Equation \zref{eq:DFT_in_DFT} reduces to Equation \zref{eq:full_energy} for the case that the active and environment regions are treated at the same level of theory \cite{Claudino2019}.

Importantly $\mathcal{E}[\gamma_{emb}^{act}] = tr(\gamma_{emb}^{act} \vb{h}_{core}) + \vb{g}(\gamma_{emb}^{act})$ and does not involve $\vb{V}_{emb}$ or $\vb{P}_{proj}^{env}$. The final term in Equation \zref{eq:DFT_in_DFT} is a first-order correction that accounts for the difference between $\vb{g}(\gamma^{act}, \gamma^{env})$ and $\vb{g}(\gamma_{emb}^{act}, \gamma^{env})$, and corrects for the fact that in general $\gamma^{act} \neq \gamma_{emb}^{act}$ \cite{ goodpaster2014accurate}.

This projection based embedding approach then allows for the active system to be treated using some wave function level of theory and therefore studied using a quantum computer. The electronic energy for this is given by \cite{Claudino2019}:

\begin{equation}
\zlabel{eq:WF_in_DFT}
\begin{aligned}
    E[\Psi_{emb}^{act} ; \gamma^{act}, \gamma^{env}] &= \bra{\Psi_{emb}^{act}} \vb{H}_{emb} \ket{\Psi_{emb}^{act}} + E[\gamma_{env}] \\ 
    &+\vb{g}(\gamma^{act}, \gamma^{env}) - tr\Big( \gamma^{act} (\vb{V}_{emb}+ \vb{P}_{proj}^{env}) \Big).
\end{aligned}
\end{equation}

Importantly $\vb{H}_{emb} = \vb{h}_{emb} + \vb{g}(\Psi_{emb}^{act})$, where $\vb{g}(\Psi_{emb}^{act})$ is the two-electron operator for a given wave function method and $\vb{h}_{emb}$ is the embedded core Hamiltonian (Equation \zref{eq:H_emb}) which depends on $\gamma^{act}$ and $\gamma^{env}$ \cite{graham2022huzinaga}. As the embedding terms have been included in  $\vb{H}_{emb}$, the final correction term is therefore slightly different to Equation \zref{eq:DFT_in_DFT} \cite{goodpaster2014accurate}. The wave function calculation in Equation \zref{eq:WF_in_DFT} includes contributions from $(\vb{V}_{emb}+ \vb{P}_{proj}^{env})$ - similar to: $tr( \gamma_{emb}^{act} (\vb{V}_{emb}+ \vb{P}_{proj}^{env}))$. The correction therefore only requires subtracting $tr( \gamma^{act} (\vb{V}_{emb}+ \vb{P}_{proj}^{env}) )$, unlike in Equation \zref{eq:DFT_in_DFT}, where $\mathcal{E}$ does not use  $(\vb{V}_{emb}+ \vb{P}_{proj}^{env})$ to calculate the energy of the active system.

For the embedded system, truncating the virtual LMOs significantly reduces the computational cost for the embedded wave function calculation \cite{claudino2019simple}. This would reduce the quantum resources required, as each qubit represents an orbital. We leave this for future work.

\section{Methods}
\zlabel{sec:methods}
% Mention github
We studied the performance of our wave function projection based embedding method on a selected set of molecular systems. We developed a python package, Nbed, that utilizes the PySCF and Openfermion quantum chemistry packages to build each embedded model \cite{Sun2017, McClean2017}. The package outputs a qubit Hamiltonian for the wave function portion of an embedded problem and the classical energy corrections from density functional theory. This is freely available for use on GitHub \cite{Nbed}.

For all calculations presented, the minimal \textit{STO-3G} basis set was employed. Each global DFT calculation performed, prior to orbital localisation, used the $B3LYP$ functional. The Intrinsic Bonding Orbitals (IBO) or Subsystem Projected AO DEcomposition (SPADE) localisation procedures are used in order to isolate the molecular orbitals to the active and environment subsystem from pre-selected active atoms \cite{knizia2013intrinsic, Claudino2019}. A threshold of $95\%$ was used to select the active region when IBO was employed. This paper's Supporting Information goes into further detail on each localisation strategy.  We performed both the $\mu$-shift and Huzinaga methods for each. A Hartree-Fock calculation for the active system, using the modified core Hamiltonian, was performed for each molecular system. The second quantized molecular Hamiltonian was then constructed with Openfermion and converted to a qubit Hamiltonian using the in-built Jordan-Wigner transformation \cite{jordan10uber}. Post Hartree-Fock methods were performed with PySCF. The frozen core approximation is never used and all virtual orbitals were included in the wave function calculations. Only the occupied environment orbitals were removed from the wave function calculations of the active systems.

For the single point electronic structure calculations, each result is compared to full system CCSD(T) calculation. Each molecular geometry was obtained from PubChem \cite{Kim2021}. The potential energy surface for \ce{OH} bond stretching in water was compared to a full configuration interaction (FCI) calculation at each step, where the embedded molecular Hamiltonian at each geometry was diagonalized to find the ground state energy of the active system.

\section{Results and Discussion}

\begin{figure}[t]
     \centering
     \begin{subfigure}[b]{0.3\textwidth}
         \centering
         \includegraphics[width=0.65\textwidth]{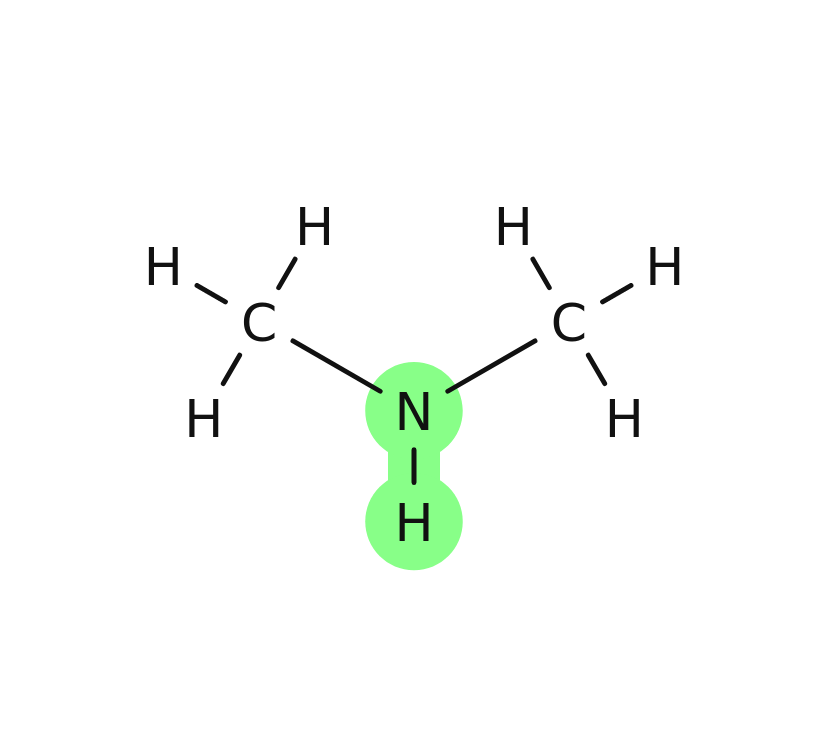}
         \caption{\ce{(CH3)2NH}}
         \zlabel{fig:N_methylmethanamine}
     \end{subfigure}
     \hfill
     \begin{subfigure}[b]{0.3\textwidth}
         \centering
    \includegraphics[width=0.6\textwidth]{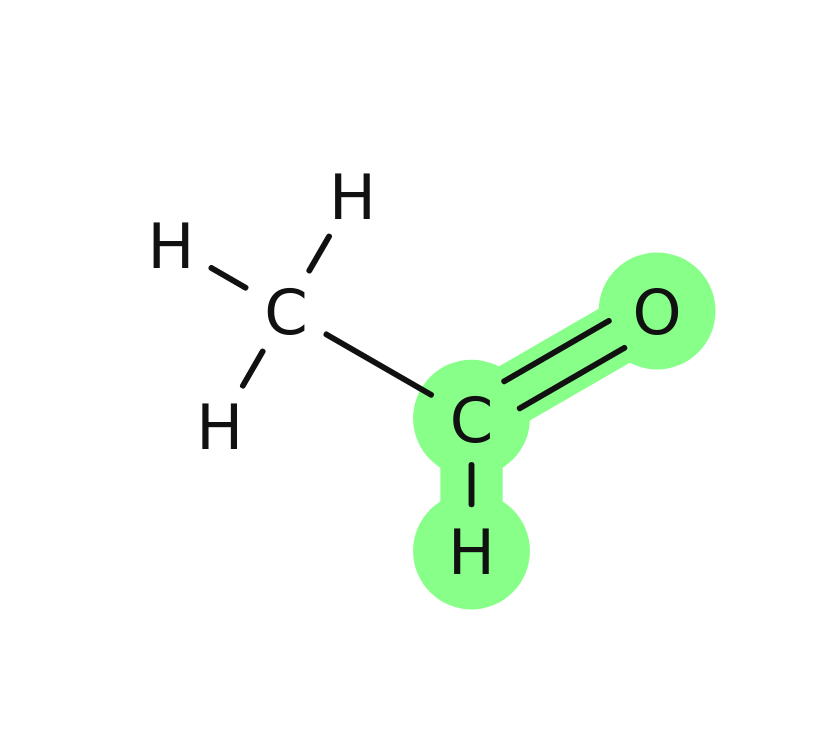}
         \caption{\ce{CH3CHO}}
         \zlabel{fig:acetaldehyde}
     \end{subfigure}
     \hfill
     \begin{subfigure}[b]{0.3\textwidth}
         \centering
         \includegraphics[width=0.6\textwidth]{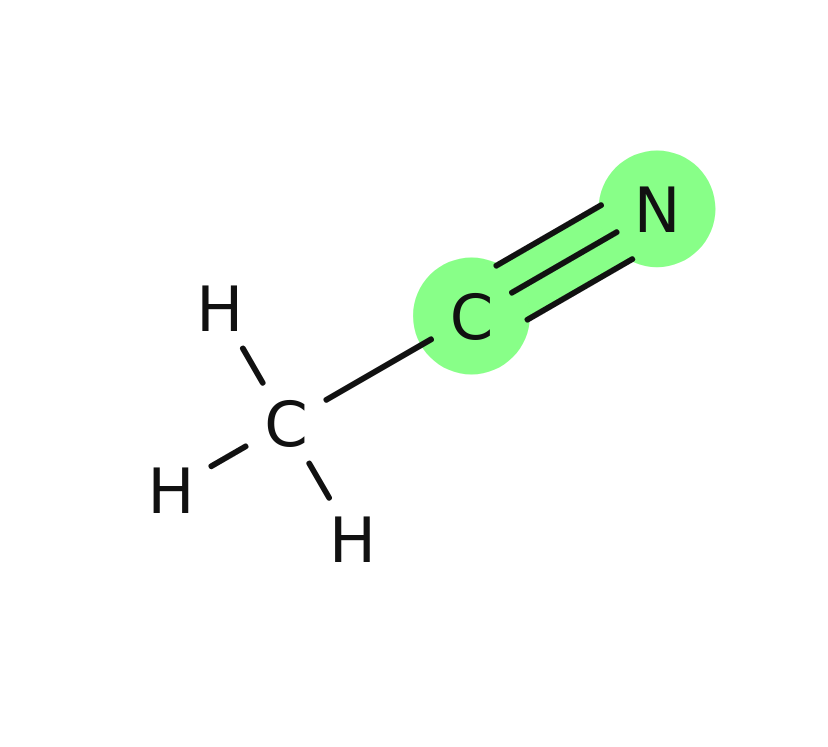}
         \caption{\ce{CH3CN}}
         \zlabel{fig:acetonitryl}
     \end{subfigure}
      \begin{subfigure}[b]{0.3\textwidth}
         \centering
         \includegraphics[width=0.65\textwidth]{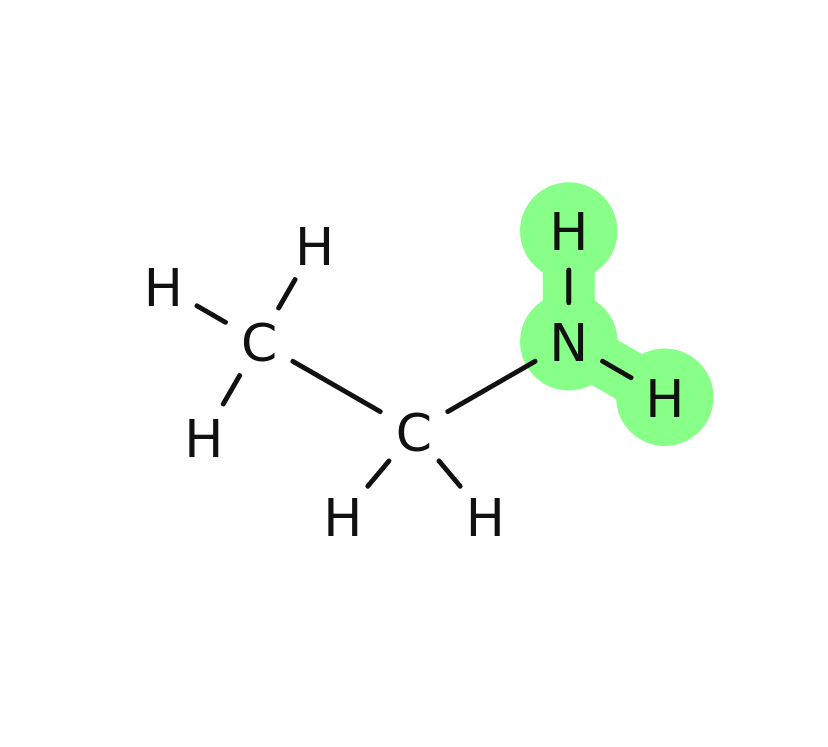}
         \caption{\ce{CH3CH2NH2}}
         \zlabel{fig:ethanamine}
     \end{subfigure}
     \hfill
     \begin{subfigure}[b]{0.3\textwidth}
         \centering
    \includegraphics[width=0.65\textwidth]{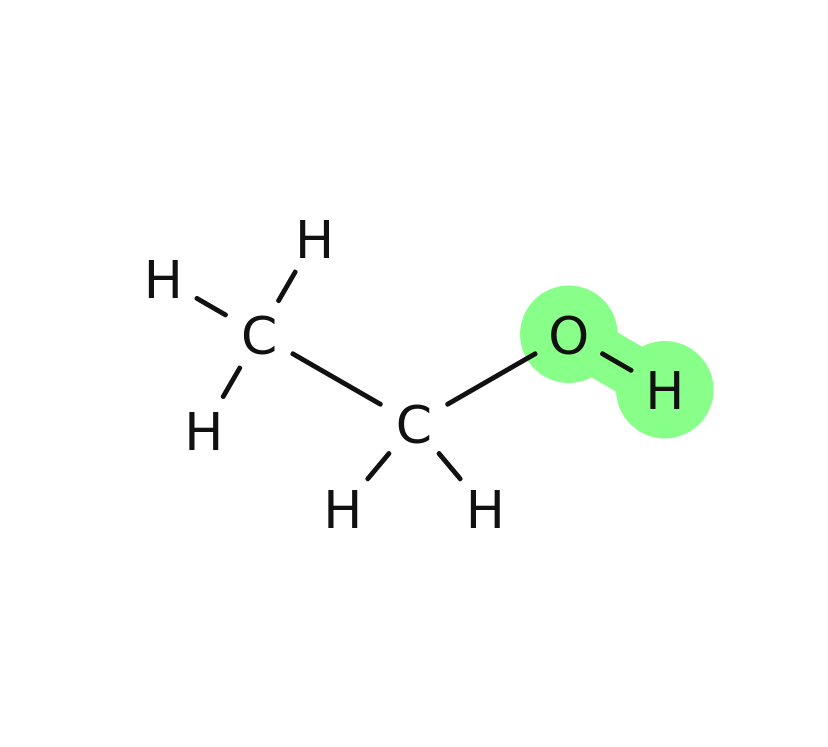}
         \caption{\ce{CH3CH2OH}}
         \zlabel{fig:ethanol}
     \end{subfigure}
     \hfill
     \begin{subfigure}[b]{0.3\textwidth}
         \centering
         \includegraphics[width=0.65\textwidth]{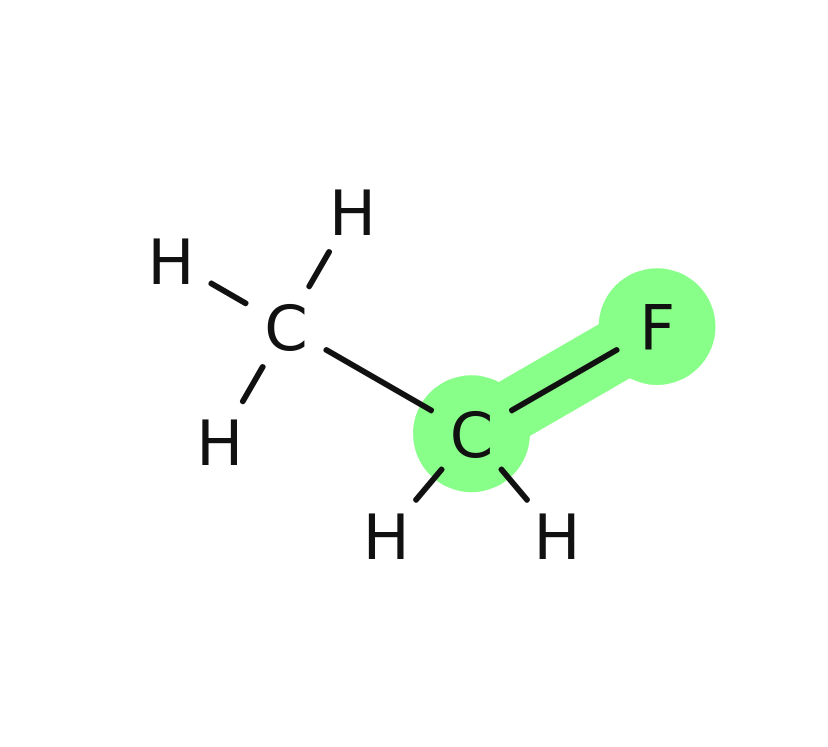}
         \caption{\ce{CH3CH2F}}
         \zlabel{fig:flouroethane}
     \end{subfigure}
   \begin{subfigure}[b]{0.3\textwidth}
         \centering
         \includegraphics[width=0.6\textwidth]{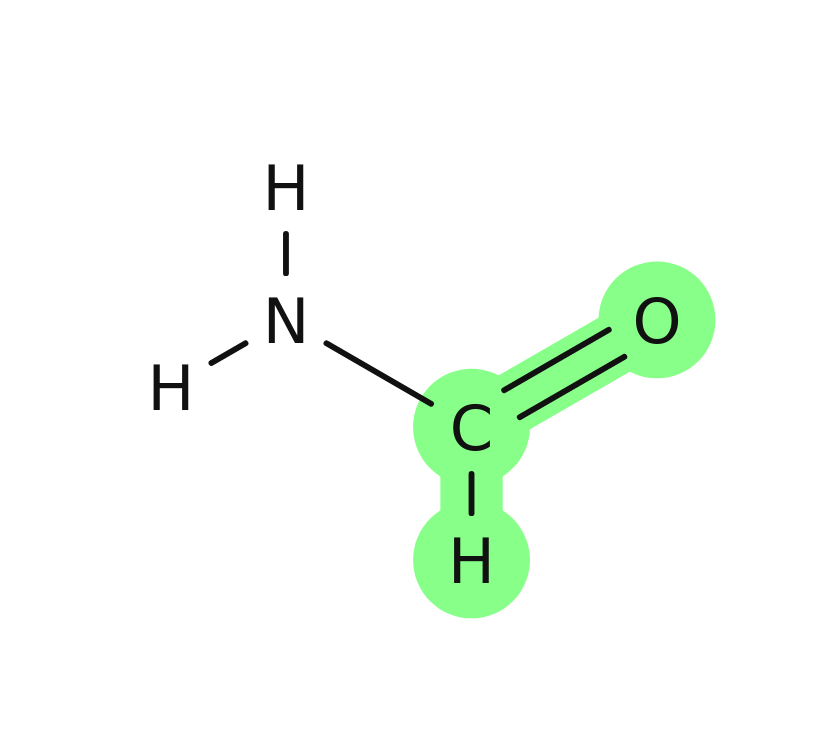}
         \caption{\ce{HCONH2}}
         \zlabel{fig:formamide}
     \end{subfigure}
     \hfill
     \begin{subfigure}[b]{0.3\textwidth}
         \centering
    \includegraphics[width=0.5\textwidth]{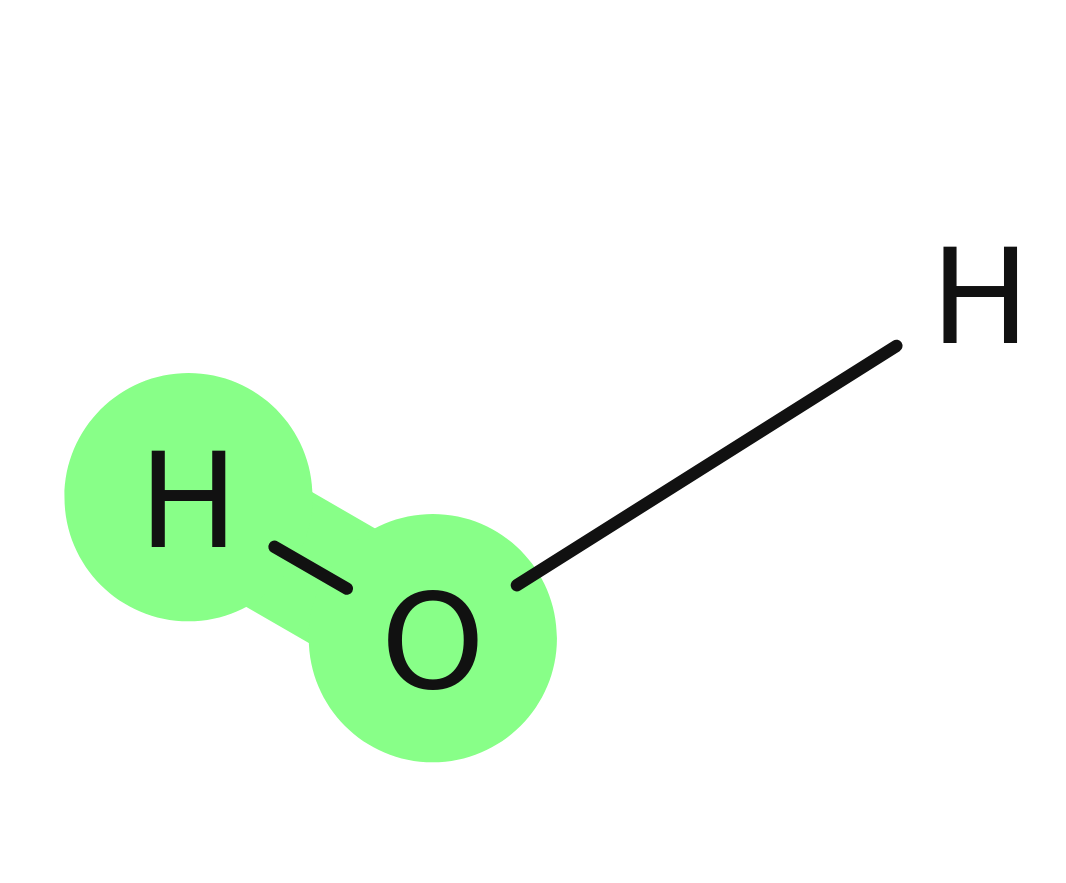}
         \caption{\ce{H2O}}
         \zlabel{fig:water_short}
     \end{subfigure}
     \hfill
     \begin{subfigure}[b]{0.3\textwidth}
         \centering
         \includegraphics[width=0.5\textwidth]{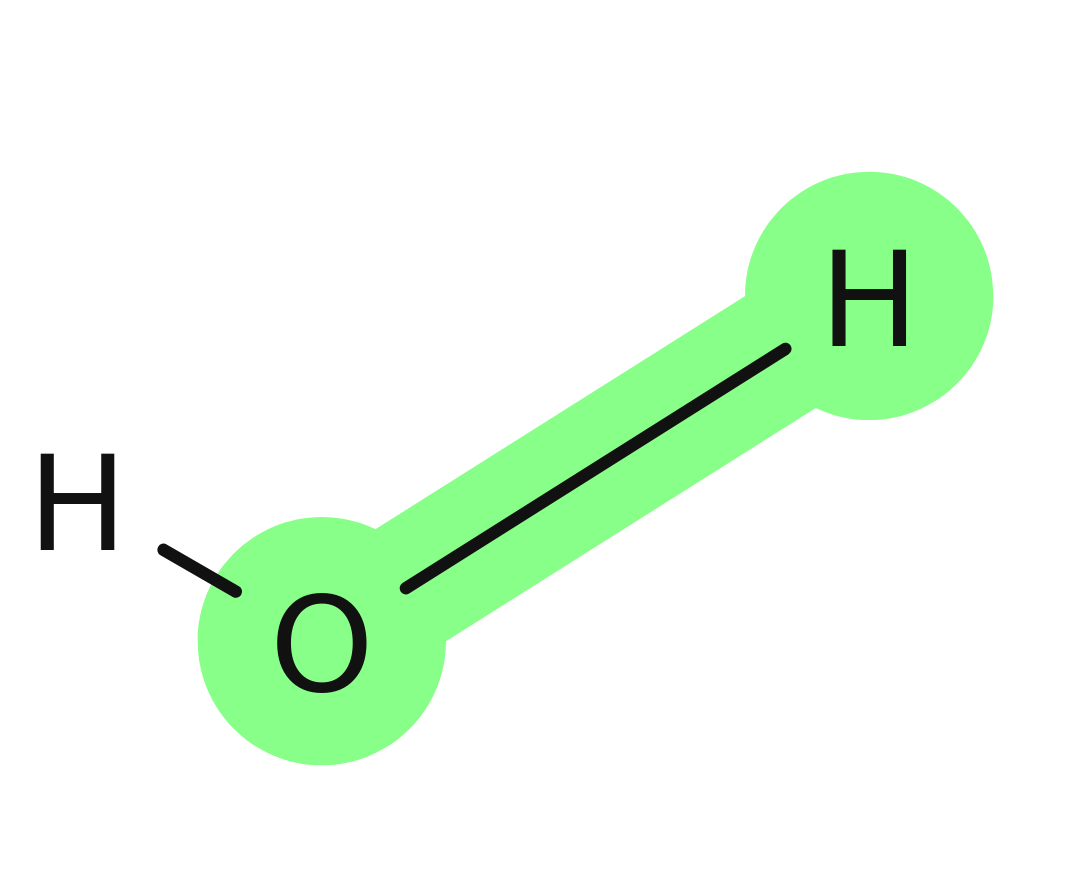}
         \caption{\ce{H2O}}
         \zlabel{fig:water_long}
     \end{subfigure}
        \caption{Planar representations of the molecules used in embedding calculations. Atoms shaded in green were selected as active for localisation procedures. Images were generated using MolView \cite{bergwerf_2022}. \zref{fig:N_methylmethanamine} N-methylmethanamine; \zref{fig:acetaldehyde} acetaldehyde; \zref{fig:acetonitryl} acetonitrile; \zref{fig:ethanamine} ethanamine; \zref{fig:ethanol} ethanol; \zref{fig:flouroethane} flouroethane; \zref{fig:formamide} formamide; \zref{fig:water_short} water (fixed bond active); \zref{fig:water_long} water (stretching bond active)}
        \zlabel{fig:active}
\end{figure}

\subsection{Molecular Ground State Energy}

\begin{figure}[h!]
    \centering
    \includegraphics[width=0.75\textwidth]{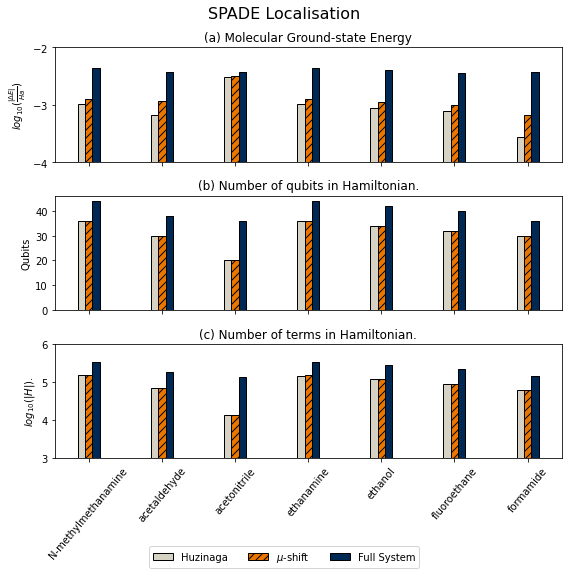}
    \caption{Results for embedding of small molecules (Figure \zref{fig:active}) using the SPADE localisation method. (a) Ground state energies for small molecules, with full-system DFT energy as reference, $\mu$-shift CCSD embedding energy in orange and Huzinaga CCSD embedding in grey. All values are given as a difference from whole system CCSD(T) energy. (b) The number of qubits needed to describe the embedded Hamiltonian, with reference showing the number required for the full system Hamiltonian. (c) The number of terms in the Jordan-Wigner encoded qubit Hamiltonian for each molecule. Again the reference gives the number needed for the full system Hamiltonian.}
    \zlabel{fig:small mols spade}
\end{figure}

Results for embedding calculations of molecular ground state energies of small molecules are shown in Figure \zref{fig:small mols spade}, with numerical values available in this paper's Supporting Information. The results for the same calculations using IBO localized orbitals can also be found in the Supporting Information. Figure \zref{fig:active} shows the partition of each molecule into active and environment orbitals which underlies our localisation methods.

Our results show increased accuracy in calculated molecular ground state energies. The Hamiltonians output using both localisation methods are reduced in size significantly; however, they still exceed the limit of what is practical to exactly solve using classical computers. Reported energies were calculated using CCSD(T)-in-DFT embedding to illustrate the application of this method to larger molecules than would be feasible using current quantum processors. As hardware continues to develop, implementation of our algorithm will be able to furnish novel results.
% Do a direct comparison of other classical - in - classical ?
Typically, results for the $\mu$-shift and Huzinaga projectors are very similar, however; the Huzinaga projector usually produces more accurate energies, in line with previous research \cite{Chulhai2017}. The number of terms in the Jordan-Wigner encoded qubit Hamiltonian, $|H|$, is typically very similar between the two projection methods.\\

In comparing the two localisation methods, we find that for acetonitrile and formamide,  SPADE and IBO partition the active system in a similar way. This results in a similar number of active MOs and hence the ground state estimation and resource requirements are very similar for these systems. For the majority of the molecules we study, SPADE includes more MOs, resulting in significantly more accurate ground state energies while still reducing the size of the Hamiltonian. However, by reducing the threshold of assigning the localized MOs from IBO to the active region, additional MOs could be included giving a similar result. See the Supporting Information for further details.

\subsection{Strong Correlation}
The impact of active region selection is demonstrated by our results in Figure \zref{fig:water}. We consider the bond dissociation of an \ce{OH} bond in water - where at high bond lengths a correlated state is created. We perform projection based embedding calculations, at different molecular geometries, for two different active regions. One has the atoms in the fixed \ce{OH} bond set active and the other has the atoms in the changing \ce{OH} bond set active. We show this pictorially in Fig. \zref{fig:water_short} and \zref{fig:water_long}.

At near equilibrium bond lengths, we see a similar performance between the different active systems (Figure \zref{fig:water}). This is due to the symmetrical structure of \ce{H2O}, hence at low bond lengths there is little difference between the two active regions. In fact, the third data point gives results for the scenario where both \ce{OH} bonds are the same length and consequently is why the results for the different active regions are the same here. However, in the correlated regime - at large bond lengths - selecting the active region to encompass the stretched atoms leads to significant improvements in energy calculation over DFT alone. This is due to correlation being effectively captured in the wave function calculation. In contrast, the full DFT calculation is plagued by deficiencies of current approximate exchange-correlation functionals \cite{cohen2008insights, cohen2012challenges}. We see in Figure \zref{fig:water} that the global DFT calculation overestimates the bond dissociation energy. This problem is attributed to static correlation \cite{cohen2008insights}. As there is no systematic way to improve the approximate exchange-correlation functionals, the way forward to describe such systems may be hybrid quantum-classical embedding. Here quantum processors could be exploited most effectively by application to only those regions of a molecule that are highly correlated.

\begin{figure}[t]
    \centering
    \includegraphics[width=0.75\textwidth]{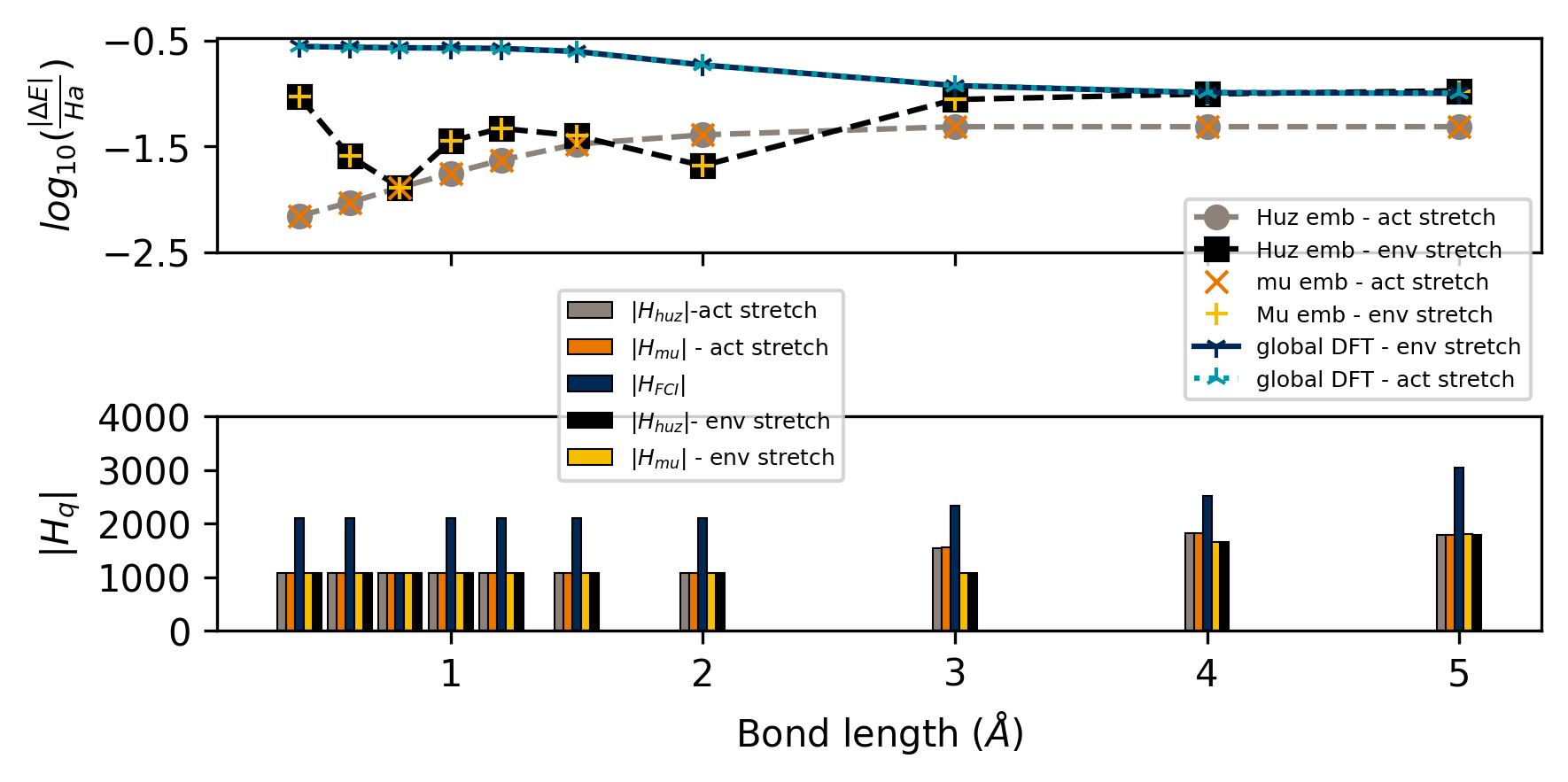}
    \caption{Potential energy curve for \ce{H2O}, with changing \ce{OH} bond length. \emph{Active stretch} result has  changing \ce{OH} bond as active region and \emph{environment stretch} result has fixed \ce{OH} bond selected as the active region. These results use SPADE localization. For each data the full problem is reduced from $14$ to $12$ qubits,  with the number of active MOs being $4$, in all cases. (top) Shows the $log$ base $10$ error with respect to the exact $FCI$ energy. (bottom) The number of terms in the Jordan-Wigner encoded qubit Hamiltonian obtained with each method. Numerical details are available in this paper's Supporting Information.}
    \zlabel{fig:water}
\end{figure}

\subsection{MO localisation method}

For the work we have presented, we only use the SPADE and IBO localized molecular orbitals. The motivation for using SPADE is primarily that it does not require a parameterised heuristic to determine the active and environment subsystems. In the IBO approach, we calculate the percentage of the $i_{th}$ LMO over atoms a user defines as the active subsystem. Any LMO that has a percentage higher than $95$ \% we assign to the active region. The SPADE approach does not require this threshold hyper-parameter. However, it does use a function of the molecular orbital coefficient matrix. Further details on both approaches are given in the Supporting Information.

The IBOs were used as they only depend on the intrinsic atomic orbital charges, rather than Mulliken charges which change erratically depending on the basis set used \cite{knizia2013intrinsic}. IBOs are therefore always well-defined, whereas other localization methods - such as Pipek-Mezey orbitals \cite{pipek1989fast}, which depend on the Mulliken charges \cite{mulliken1955electronic} - are unphysically tied to the basis set used \cite{knizia2013intrinsic}.

The effect of different localisation methods for this embedding method \cite{pipek1989fast, foster1960canonical, edmiston1963localized, hoyvik2012orbital} would be an interesting area to explore. Our software package Nbed can run any method given by PySCF, and users can also build their own localisation strategies themselves. 

\section{Conclusion}

We have used the projection-based embedding technique \cite{Manby2012} to reduce the size of an electronic structure calculation studied at the wave function level. The molecular problem is split into active and environment parts, each solved using different levels of theory. The active part is treated using a wave function approach and an embedded qubit Hamiltonian is generated. Solving this provides $E_{act}^{WF}= \bra{\Psi_{emb}^{act}} \vb{H}_{emb} \ket{\Psi_{emb}^{act}}$. The whole system and environment are treated using density functional theory and the overall electronic energy is found as via an additive procedure \cite{goodpaster2014accurate, Chulhai2017, graham2022huzinaga}. This is similar to the own n-layered integrated molecular orbital and molecular mechanics (ONIOM) subtractive framework \cite{svensson1996oniom}. What is included in the active region can be modified and thus the size of the quantum problem varied. This allows users to tune their problem to available hardware.

For a small collection of molecules, too large for whole system quantum simulation and classical FCI, we have shown that this method produces more accurate energies than full system DFT when each is compared to full system CCSD(T). Furthermore, we have shown its ability to capture the effects of strong correlation by investigating the bond dissociation of \ce{H2O}.

As this approach generates an embedded qubit Hamiltonian, it is agnostic to the quantum algorithm used to solve $\vb{H}_{emb}$. NISQ friendly approaches such as the VQE algorithm can therefore be used, but also fault-tolerant methods such as quantum phase estimation (QPE) \cite{aspuru2005simulated}.

Moreover, as our method outputs a qubit Hamiltonian, different resource reduction techniques can be used in conjunction with it; for example, the contextual-subspace approach of Kirby \textit{et al}, \cite{Kirby2020} or the entanglement forging approach of Eddins \cite{eddins2022doubling}. Similarly, the $\mathcal{Z}_{2}$-symmetries of the problem can also be removed via qubit tapering \cite{Bravyi2017}.

As our method does not rely on imposing constraints on the system studied or costly parameter fitting, it may be reasonably combined with other hybridisation techniques which do\cite{Stenger2021, Rossmannek2020}.

Further work is planned to develop this method. As significant resource reduction is achieved by localisation of only the occupied orbitals, virtual orbital localisation could lead to a greater reduction in computational resources \cite{claudino2019simple}. 

We anticipate that our code will allow researchers to study molecules of real chemical interest on quantum computers. We welcome readers to make use of this, which is freely available on Github \cite{Nbed}.

\begin{acknowledgement}
A. R. and M. I. W. acknowledge  support from the Engineering and Physical Sciences Research Council (EPSRC) (EP/L015242/1 and EP/S021582/1 respectively). M.I.W. also acknowledges support from CBKSciCon Ltd. P. V. C. is grateful for funding from the European Commission for VECMA (800925) and EPSRC for SEAVEA (EP/W007711/1). We would like to thank Prof. Dieter  Kranzlm\"uller at the Leibniz Supercomputing Centre (LRZ), who provided access to their ATOS Quantum Learning Machine simulator for some of the computations. The authors would also like to thank Dr. David A. Herrera-Mart{\'i} for useful preliminary discussions on embedding. A.R. and M.I.W. contributed equally to this work.
\end{acknowledgement}

\begin{suppinfo}
Further details on the localization methods employed, active atom selection, embedded self-consistent field implementation and numerical results are supplied.
\end{suppinfo}

% \end{multicols}
\bibliography{references}
\end{document}